\documentclass[final,3p,twocolumn]{elsarticle}
\usepackage{graphicx}
\usepackage{amssymb}
\usepackage{amsmath}
\biboptions{numbers,sort&compress}
\journal{Physics Letters B}
\begin{document}

\begin{frontmatter}

\title{Anti-bubble effect of temperature \& deformation: a systematic study for nuclei across all mass regions between A $=$ 20$-$300}
\author[a,b]{G. Saxena}
\author[a,b]{M. Kumawat}
\author[c]{Mamta Aggarwal\corref{cor1}}
\cortext[cor1]{corresponding authors: Mamta Aggarwal, mamta.a4@gmail.com}
\address[a]{Department of Physics, Government Women Engineering College, Ajmer-305002, India}
\address[b]{Department of Physics, School of Basic Sciences, Manipal University, Jaipur-303007, India}
\address[c]{Department of Physics, University of Mumbai, Kalina Campus, Mumbai-400098, India}

\begin{abstract}
Temperature dependent relativistic mean-field (RMF) plus BCS approach has been used for the first time to investigate the anti-bubble effect of the temperature and deformation in the light, medium-heavy and superheavy nuclei. Influence of temperature  is studied on density distribution, charge form-factor, single particle (s.p.) energies, occupancy, deformation and the depletion fraction (DF). At T $=$ 0, the quenching effect of deformation is predominant. DF is found usually less in oblate deformation than in prolate. DF decreases with increasing prolate deformation even though the 2s-orbit is empty which shows the role of deformation in central depletion apart from the unoccupancy in s-orbit as is usually believed. As T increases, the occupancy of s-orbit increases, shell structure melts, the deformation vanishes and the weakening of central depletion is solely due to the temperature. The bubble effect is eliminated at T $\approx$ 3$-$5 MeV as indicated by DF and the charge form factor. The temperature effect is found less prominent in superheavy bubble nuclei where the role of shell effects is indicated.
\end{abstract}


\end{frontmatter}

The  phenomenon of bubble structure characterized by the unconventional depletion of the central density of nucleons is becoming a thrust area of nuclear physics research on the experimental as well as theoretical fronts~\cite{nature,campi,todd,grasso,khan,wang,wang1,grasso1,yao1}. The bubble effect is related to the shell effects associated with the occupancy in zero angular momentum (s) orbitals that have large central density and cause central density depletion, if empty. The non-zero $\ell$ orbitals being suppressed do not contribute to the central density. However, the unoccupied s-orbit and nearby single-particle shells can favor collective correlations and thus lower or even wash out the central density depletion. Thus the maximum bubble effect comes by the s-orbital surrounded by larger $\ell$ orbitals well separated in energy from nearby s.p. states to ensure the weak dynamical correlations. So far the impact of shell structure on the bubble effect has been reported in nuclei below $^{208}$Pb, but the bubble phenomena in heavier systems~\cite{schuetrumpf,sobi,decharge,sksingh,ikram,bender}, which is mainly driven by combined effect of Coulomb and symmetry energy, may also have an impact of quantum shell structure but not as prominent as it is in the lighter nuclei.\par

The pairing correlations and the dynamical quadrupole shape effects have been observed~\cite{wu} to hinder the bubble effect~\cite{yao,yao1,grasso,li} but unable to eliminate it~\cite{duguet}. However the temperature (T) appears to have an anti-bubble effect that might quench or completely wash out the bubble effect at a certain critical value of T~\cite{TAN,plb2018}. We have recently shown the  effect of temperature on the occupancy of s-orbit which indicated the possible role of temperature in central depletion~\cite{plb2018}. The nuclear tensor-force and the pairing correlations are conjectured to have important implications in the shell evolution and bubble structure~\cite{nakada,khan,grasso,wu}. The influence of Coulomb  and pairing energy, neutron to proton ratio and the nuclear deformation on the central density depletion has been discussed in one of our recent work~\cite{BKbubble}. Since the inclusion of temperature in a nucleus alters the nuclear shell structure profoundly, various significant changes are induced in the intrinsic shape and deformation~\cite{MAPLB,MAPRC,MAPRC80,MAPRC69} of the nucleus. At a critical temperature, where the shell effects melt away with the shape change to spherical and deformation reducing to zero, the bubble structure being associated with shell effects is also expected to undergo transitions. So far, the temperature (T) dependence of central density depletion has not been addressed in any work except in recent communications~\cite{TAN,plb2018}. The  central depletion becomes less pronounced as T increases and disappears at a critical temperature as shown by Ref.~\cite{TAN} using the Skyrme Hartree-Fock mean-field~\cite{TAN} approach for spherical light bubble nuclei $^{34}$Si and $^{22}$O. Since the bubble phenomenon is believed to exist in all the mass regions including superheavy region, in spherical as well as in the deformed nuclei, a comprehensive study is required to examine the role of temperature and deformation on bubble effect, which is precisely the objective of this work.\par

Here we present a theoretical study on the effect of temperature (T) on the spherical and deformed bubble nuclei. The quenching effect of deformation at T $=$ 0 and the interesting interplay between the deformation and temperature on the bubble effect  at T$>$ 0 has been investigated. For this purpose, we have  used the temperature dependent relativistic mean-field (RMF) plus state dependent Bardeen-Cooper-Schrieffer (BCS) theory~\cite{saxena,plb2018,BKbubble,saxena3}. We evaluate density and depletion fraction (DF $=$ ($\rho_{max}-\rho_{c})/\rho_{max}$, where $\rho_{max}$ and $\rho_{c}$ are maximum and central densities), charge form-factor and the occupancy as a function of temperature for potential bubble candidates $^{34}$Si, $^{46}$Ar, $^{22}$O, $^{34}$Ca in light region, $^{294}$Og, $^{302}$Og,~\cite{schuetrumpf} and $^{292}120$~\cite{li} in superheavy region ~\cite{plb2018,nature,duguet,todd,grasso,khan,wang,wang1,grasso1,yao1,li,schuetrumpf,wu} and few new bubble candidates  $^{22}$Si, $^{56}$S, $^{58}$Ar, $^{184}$Ce predicted by us~\cite{plb2018}. In RMF approach, apart from TMA parameter~\cite{suga}, we also use NL3$^{*}$~\cite{Lalazissis09} and DD-ME2~\cite{Lalazissis05} parameters for comparison. \par

RMF calculations have been carried out using the model Lagrangian density with nonlinear terms for both the ${\sigma}$ and ${\omega}$ mesons along with TMA parametrization described in Refs.$~$\cite{saxena,plb2018,saxena3,suga}. The corresponding Dirac equations for nucleons and Klein-Gordon equations for mesons obtained with the mean-field approximation are solved by the expansion method on widely used axially deformed Harmonic-Oscillator basis \cite{gambhir,geng1}. The basis deformation $\beta_0$ is set equal to $\beta_{2m}$ and the quadrupole constrained calculations are performed \cite{hirata}, in order to obtain the potential energy surfaces (PESs) and the ground-state deformations \cite{geng1}. We have used the expansion in 12 (for light nuclei) and 20 (for superheavy nuclei) oscillator shells for both the fermion and boson fields along with $\hbar \omega_{0} = 41A^{-1/3}$ for fermions. The convergence of the calculations has been tested. \par

For our calculations, we use a delta force V = -V$_0 \delta(r)$ with the strength V$_0$ = 350 MeV fm$^3$  for pairing interaction, which has been used for the successful description of drip-line nuclei \cite{saxena3,yadav2004,saxena5} and bubble nuclei \cite{plb2018,saxena}. Whenever the zero-range $\delta$ force is used either in the BCS or in the Bogoliubov framework, a cutoff procedure must be applied i.e. the space of the single-particle states, where pairing interaction is active, must be truncated. This is not only to simplify the numerical calculation but also to simulate the finite-range (more precisely, long-range) nature of the pairing interaction in a phenomenological way \cite{Dobaczewski1995,Goriely2002}. In the present work, the single-particle states subject to the pairing interaction are confined to the region satisfying
\begin{equation}
\epsilon_i-\lambda\le E_\mathrm{cut},
 \end{equation}
 where $\epsilon_i$ is the single-particle energy, $\lambda$ is
 the Fermi energy, and $E_\mathrm{cut} = 8.0$ MeV. For further details of these formulations, one may refer $~$\cite{gambhir,geng1,saxena4,saxena5}.\par

To incorporate the temperature (T) degree of freedom in our RMF formalism~\cite{gambhir,geng1,saxena4}, we calculate the occupation probabilities $v^2_{j}$ in the formula of particle number condition (${\sum_{j}} (2j+1)\, v^2_{j}\,=\,N $) using the following equation
\begin{eqnarray}
v^2_{j}\,=\,\frac{1}{2}\bigg(1 -
\frac{\varepsilon_j\,-\,\lambda}{\tilde{\varepsilon}_j}[1\,-\,2f(\tilde{\varepsilon}_j, T)]\bigg)
\end{eqnarray}
with
\begin{eqnarray}
f(\tilde{\varepsilon}_j, T)\,=\,\frac{1}{(1\,+\,exp[\tilde{\varepsilon}_j/T])}
\end{eqnarray}
and
\begin{eqnarray}
\tilde{\varepsilon}_j\,=\,\sqrt{\big(\varepsilon_j\,-\,\lambda
\big)^2\,+\,{\Delta_{j}^2}}
\end{eqnarray}
The function $f(\tilde{\varepsilon}_j, T)$ represents the Fermi Dirac distribution function for quasi particle energies $\tilde{\varepsilon}_j$. The readers may refer to Refs.$~$ \cite{tbrack,tquentin} for temperature dependent non-relativistic approach, and Refs.$~$\cite{tgambhir,tagrawal,ttapas,tsenthil,tbharat} for the temperature dependent RMF. \par

It is usually believed that the temperature dependent BCS leads to a sharp pairing collapse at T around 0.5 $-$ 0.6 MeV indicating the transition from the superfluid to normal phase \cite{tagrawal,goodman}. However, various approximations have already shown that the thermal fluctuations in finite systems as atomic nuclei smooth out this phase transition, resulting in a thermal gap, which monotonically decreases with increasing T (see, e.g., Refs. \cite{TAN,goodman,quang,dinh,moretto,dinh1,dinh2}). This agrees with our preliminary calculations on pairing effect using the state-dependent BCS method. The pairing does not collapse at T $\approx$0.5 MeV, which shows the reliability of our formalism for temperature dependent calculations.\par

\begin{figure}[htb]
\centering
\includegraphics[width=8cm,height=10cm,keepaspectratio]{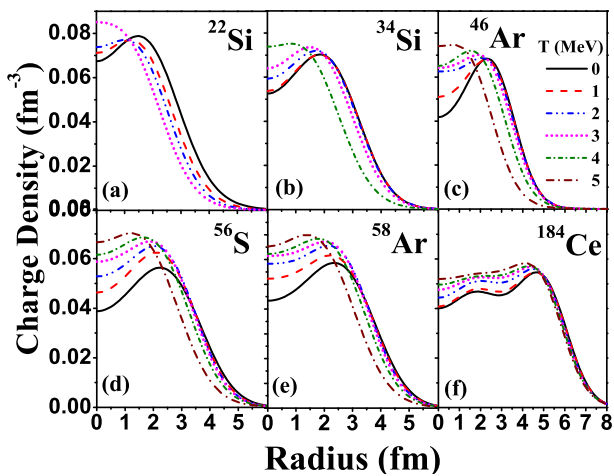}
\caption{(Colour online) Charge density vs. radius for different temperature (T).}
\label{fig1}
\end{figure}

In Fig. \ref{fig1}, we show the charge density of bubble nuclei $^{22}$Si, $^{34}$Si, $^{46}$Ar, $^{56}$S, $^{58}$Ar and $^{184}$Ce as a function of radius for temperature varying from T $=$ 0 to 5 MeV. At T $=$ 0, all the nuclei show significant central density depletion. As T increases, the central depletion decreases and eventually vanishes at a critical temperature T$_c$ $\approx$ 3$-$5 MeV.  The critical temperature T$_c$ is different for the different nuclei. In  medium-heavy nucleus, $^{184}$Ce, the central depletion does not vanish completely even at T$=$ 5 MeV which means the T$_c$ for $^{184}$Ce is beyond 5 MeV. T$_c$ in $^{34}$Si is 4 MeV in agreement with that predicted by Ref.~\cite{TAN}.  \par

\begin{figure}[htb]
\centering
\includegraphics[width=8cm,height=10cm,keepaspectratio]{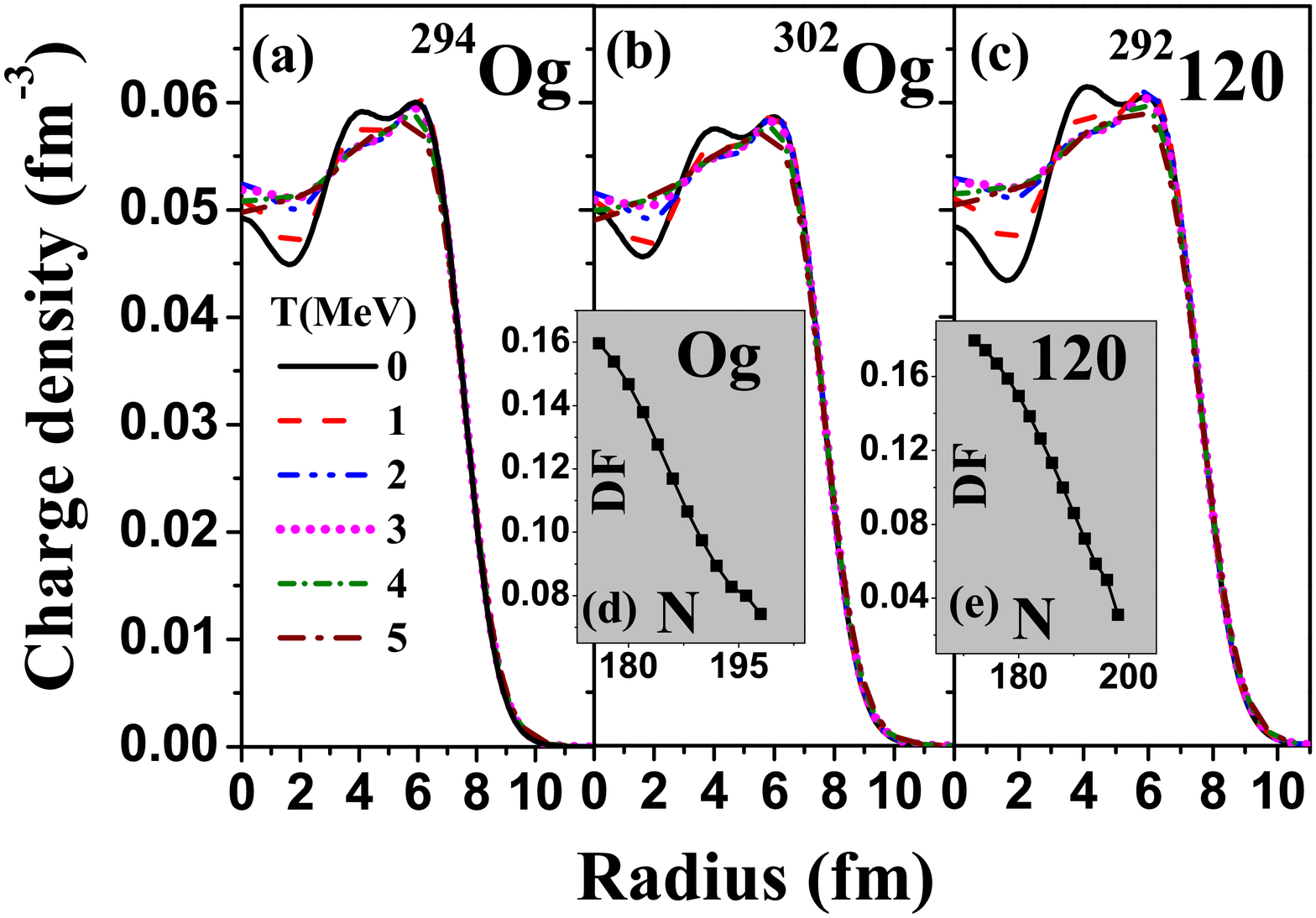}
\caption{(Colour online) Charge density of $^{294}$Og, $^{302}$Og and $^{292}120$ vs. radius at different T. Insets show DF vs. N at T=0 for (d) Og isotopes (e) Z$=$120 isotopes.}
\label{fig2}
\end{figure}

\begin{figure}[htb]
\centering
\includegraphics[width=8cm,height=10cm,keepaspectratio]{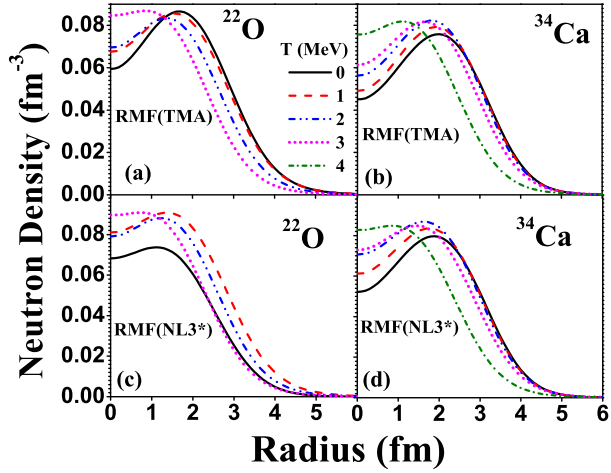}
\caption{(Colour online) Neutron density of $^{22}$O and $^{34}$Ca vs. radius at different temperature (T).}
\label{fig3}
\end{figure}

Extending our calculation on temperature dependence of bubble nuclei to the superheavy region, we evaluate the charge density of the bubble nuclei $^{294}$Og, $^{302}$Og~\cite{schuetrumpf} and $^{292}120$~\cite{li} shown in Fig. \ref{fig2}. It is observed that the effect of temperature on the central depletion of superheavy nuclei is not as significant as in the lighter nuclei. Although the bubble effect in superheavy region is driven mainly by the interplay between the Coulomb and the nuclear strong forces~\cite{schuetrumpf,plb2018}, but the shell effects also seem to play a role~\cite{decharge,bender} where not only the depopulation of the s-state is important but also the occupation of high-j states building the density in the surface region is important. In Fig. \ref{fig2}, the central depletion is decreasing with increasing temperature but not vanishing completely as it does at T$_c$ in light nuclei. The slight reduction in central depletion by the temperature could be due to washing away of the underlying shell effects which appear to play a role. As seen in the inset of Fig. \ref{fig2} (and our work~\cite{plb2018}) that for a fixed Z, DF is decreasing with increasing N which shows the role of nucleon attractive forces balancing the Coulomb forces and consequently decreasing the central depletion and hence the DF. However, with increasing isospin, the higher energy levels get occupied. The particles near the fermi level occupy further higher levels as T increases. Consequently, the central depletion decreases to some extent due to the structural changes, but, sustains due to the combined effect of large repulsive Coulomb forces and the attractive nucleon forces. This is in contrast to the lighter nuclei where the predominant shell effects get washed out due to increasing T and eliminate the bubble effect completely at T$_c$. This shows a subtle role of shell effects in central density depletion in superheavy systems as also suggested in Refs.~\cite{decharge,bender}.  \par

Fig. \ref{fig3} displays the neutron density for the case of  neutron bubble nuclei $^{22}$O and $^{34}$Ca using TMA~\cite{suga} and NL3$^{*}$~\cite{Lalazissis09} parameters of RMF that show good agreement. The reduction in the central depletion due to temperature is evident. \par
\begin{figure}[htb]
\centering
\includegraphics[width=8cm,height=10cm,keepaspectratio]{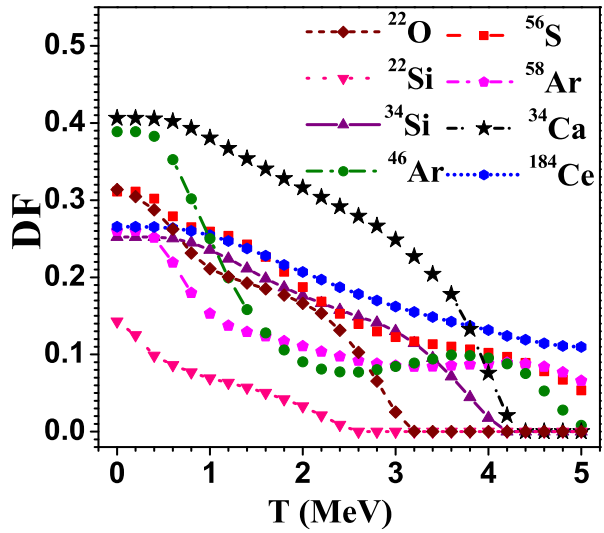}
\caption{(Colour online) Depletion fraction (DF $=$ ($\rho_{max}-\rho_{c})/\rho_{max}$, where $\rho_{max}$ and $\rho_{c}$ are maximum and central densities)  vs. T.}
\label{fig4}
\end{figure}
The depletion fraction which quantifies the central depletion, is plotted as a function of T in Fig. \ref{fig4} for all the potential bubble nuclei studied here. A gradual decrease of DF with the increasing temperature exhibits the quenching of bubble in all the bubble candidates shown. $^{34}$Ca shows the maximum DF (at T $=$ 0) showing strong neutron bubble effect. The critical temperature for $^{22}$O and $^{22}$Si for the washing out of bubble effect is around 3 MeV whereas that for $^{34}$Si is 4 MeV in agreement with Ref.~\cite{TAN}. The slope of DF vs. T appears to be relatively smaller in case of $^{184}$Ce and the critical T$_c$ value also appears to be more than 5 MeV as the DF has not vanished at 5 MeV (observed in Fig. 1 also). However, this points towards the speculation of a possible role of Coulomb repulsion in the central depletion in medium-heavy nucleus $^{184}$Ce. Among the proton bubble candidates, $^{46}$Ar shows the highest DF indicating a strong bubble nucleus. With increasing temperature, the sharp drop in DF  to a vanishingly small value between T$\approx$ 3-5 MeV shows the strong anti-bubble effect in $^{46}$Ar. However, Our calculations using the density-dependent point coupling variant of RMF model shows~\cite{BKbubble} no central depletion in $^{46}$Ar, whereas RMF model (with TMA parameter) has shown~\cite{plb2018} significant depletion due to inversion of 2s$_{1/2}$ and 1d$_{3/2}$ states without including the tensor force~\cite{todd,grasso,khan,wu}. This points towards the model dependency ~\cite{BKbubble} as well as certain uncertainties in the existence of proton bubble in $^{46}$Ar, which might hopefully be sorted out with the upcoming experimental facilities SCRIT, RIBF~\cite{suda,suda1}. Fortunately, among the best bubble candidates predicted above, the nuclei $^{46}$Ar, $^{34}$Si and $^{22}$O seem to be possible to study, in principle,  with the slow RI beams in the upcoming SCRIT facility~\cite{suda,suda1} in the near future.\par

\begin{figure}[htb]
\centering
\includegraphics[width=8cm,height=10cm,keepaspectratio]{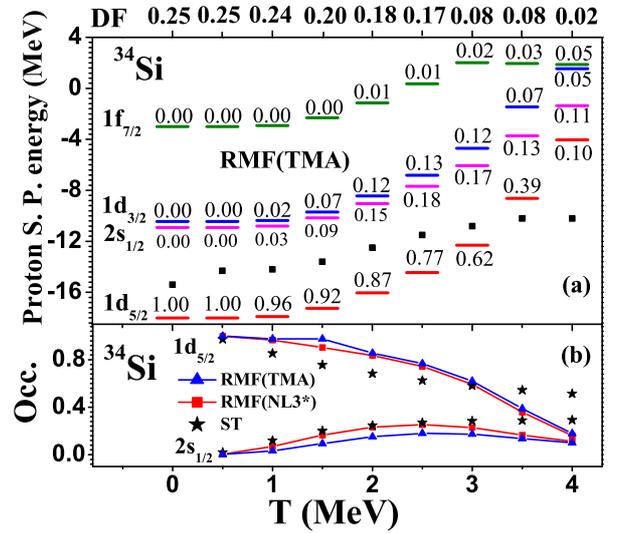}
\caption{(Colour online) (a) Proton s.p.energies of $^{34}$Si vs. T. Occupancy is marked by numbers on each level. Values of DF are also indicated on the top. (b) Occupation probabilities vs. T}
\label{fig5}
\end{figure}

\begin{figure}[htb]
\centering
\includegraphics[width=8cm,height=10cm,keepaspectratio]{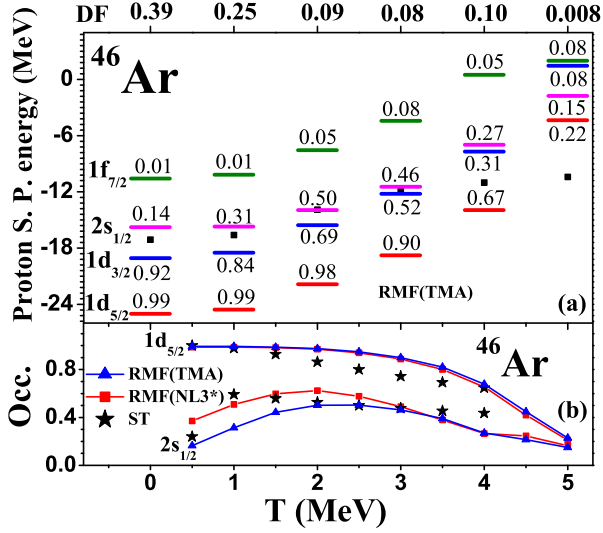}
\caption{(Colour online) Same as Fig. 5 but for $^{46}$Ar.}
\label{fig6}
\end{figure}

The dependence of temperature on the proton single particle states ($\pi$2s$_{1/2}$) is examined for $^{34}$Si and $^{46}$Ar as the representative examples due to their accessibility in future experiments~\cite{suda,suda1}. Figs. \ref{fig5}(a) and \ref{fig6}(a) show the proton single particle states 1d$_{5/2}$, 2s$_{1/2}$, 1d$_{3/2}$ and 1f$_{7/2}$ of $^{34}$Si and $^{46}$Ar respectively at temperature T $=$ 0 to 4(5) MeV. Fermi energy is also displayed by solid squares at each temperature. With increasing T, all the states become lesser and lesser bound and the rearrangement of particles near the Fermi level takes place. For T $>$ 1 MeV, the occupancy in 1d$_{5/2}$ state decreases while filling up the higher states. The filling of the particles in 2s$_{1/2}$ state reduces DF. In $^{34}$Si, DF reaches a vanishingly small value erasing the bubble effect at T $\approx$ 4 MeV. Similarly, DF almost vanishes at T $\approx$ 3-5 MeV for $^{46}$Ar indicating the anti-bubble effect of T in Fig. \ref{fig6}. Inversion of 2s$_{1/2}$ and 1d$_{3/2}$ states is visible in case of  $^{46}$Ar in Fig. \ref{fig6} (shown for T $=$ 0 in our previous work~\cite{plb2018}). At T $=$ 0, almost vacant 2s$_{1/2}$ state is lying above the fully occupied 1d$_{3/2}$ state with a large energy gap between them making $^{46}$Ar a strong bubble candidate. As T increases, the energy gap between the two states 2s$_{1/2}$ $\&$ 1d$_{3/2}$ reduces and eventually almost vanishes where both the states come closer and get occupied at T around 3 MeV. The DF reduces from a value of 0.39 at T $=$ 0 to 0.08 at T $=$ 3 MeV showing the quenching effect of T. The central depletion almost vanishes to DF $=$ 0.008 at T $=$ 5 MeV. In \ref{fig5}(b) and \ref{fig6}(b), we compare the occupation probability (occ.) of $\pi$2s$_{1/2}$ and $\pi$1d$_{5/2}$ orbits of $^{34}$Si and  $^{46}$Ar, respectively, calculated by RMF+BCS approach using TMA and NL3$^*$ parameters and Statistical theory (ST) of hot nuclei~\cite{MAPLB,MAPRC}. Variation of occ. with T computed by both the parameters of RMF and both the theories show good agreement. \par

\begin{figure}[htb]
\centering
\includegraphics[width=8cm,height=10cm,keepaspectratio]{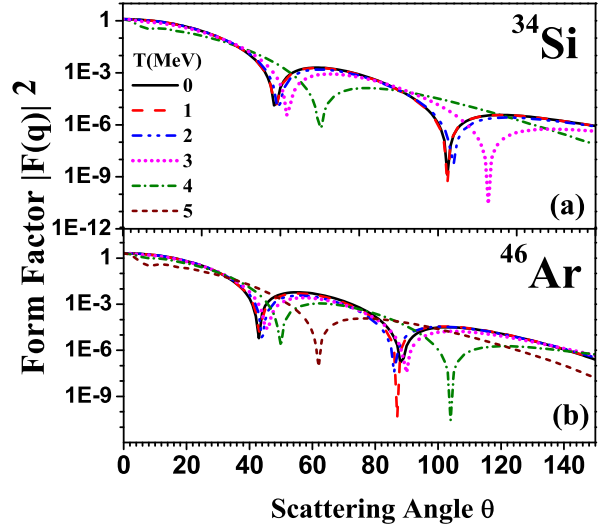}
\caption{(Colour online) Nuclear charge form factors vs. scattering angle for $^{34}$Si and $^{46}$Ar.}
\label{fig7}
\end{figure}

Our computed charge form-factor, a useful physical observable of central depletion, which can be measured through the elastic electron-nucleus scattering experiments~\cite{hofstadter,forest,donnelly}, has been shown as a function of scattering angle at different temperatures in Fig. \ref{fig7} for $^{34}$Si and $^{46}$Ar. It has been shown~\cite{meucci} that the presence of a central depletion in charge density shifts the zeros of the form factor. This was exhibited for $^{46}$Ar and $^{34}$Si by us~\cite{plb2018} at T $=$ 0, where our computed form factor had shown good agreement with that of the experiment for $^{48}$Ca. The quenching of bubble effect is evident at T $>$ 3 MeV where the peak of our computed form factors shift and the angular distribution of bubble nuclei (for T $=$ 0 to 3 MeV) and non-bubble nuclei (for T $>$ 3 MeV) are different. While T increases from 0 to 3 MeV, the small variations in the charge density are not visible in the form factor distribution, whereas for T $>$ 3 MeV, the quenching of bubble is evident in the distinctly separated peaks of the form factor.  Interestingly, in the case of $^{34}$Si, the second peak of the angular distribution of form factor goes completely out of phase and is no longer observed in the range of scattering angle shown. For $^{46}$Ar, a non-bubble is predicted  by the distinct peak of the charge form factor at T $=$ 5 MeV which agrees our prediction of DF vanishing at T $=$ 5 MeV as seen in Fig. 4. The identification of a bubble and a non-bubble by the angular distribution of charge form factor seen in Fig. \ref{fig7} calls for more experimental data to enable identification of bubble nuclei.\par

There have been indications that apart from temperature, the nuclear deformation has quenching effect ~\cite{duguet,khan,grasso1,yao1,wu} on the bubble structure. Since the temperature is known to (i) erase the bubble effect (as shown in the present work and ~\cite{TAN}) and (ii) wash out the shell effects and drive the nucleus towards sphericity with zero deformation (as shown by us using the statistical theory ~\cite{MAPLB,MAPRC80,MAPRC69}), it would be interesting to systematically study the central depletion under the influence of deformation and temperature, together. For this, we have picked up well deformed potential bubble nuclei $^{40}$Mg and $^{44}$S identified in our recent work~\cite{BKbubble}, and $^{24}$Ne and $^{32}$Ar reported recently by Shukla \textit{et al.}~\cite{shukla}. \par

\begin{figure}[htb]
\centering
\includegraphics[width=8cm,height=10cm,keepaspectratio]{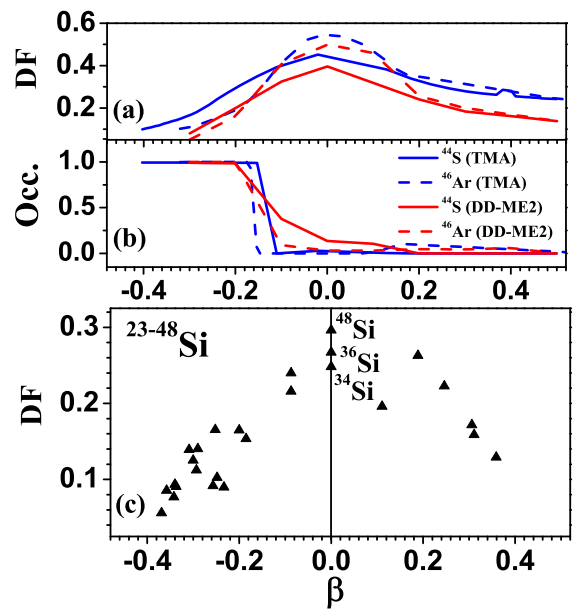}
\caption{(Colour online) (a) DF of $^{44}$S and $^{46}$Ar (b) Occ. of $\pi$2s$_{1/2}$ state and (c) DF of Si isotopes vs. deformation parameter $\beta$.}
\label{fig8}
\end{figure}

In Fig. \ref{fig8}, we plot (a) the depletion fraction (DF) and (b) the occupancy of the 2s$_{1/2}$ states of protons for deformed $^{44}$S and a neighbouring spherical nucleus $^{46}$Ar as a function of quadrupole deformation parameter $\beta$ $=$ 0.4 to -0.4 at T $=$ 0. Here we have used TMA~\cite{suga} and DD-ME2~\cite{Lalazissis05} parameters for the calculations which show a very good agreement for $^{44}$S and $^{46}$Ar. Table I shows the $\beta$ values calculated by RMF and Nilson Strutinsky Method (NSM)~\cite{MAMPRC89} that show reasonable agreement with the available experimental~\cite{nndc} and other theoretical values~\cite{Moller,Goriely,Horiuchi}. \par

In Fig. \ref{fig8}(a), DF is maximum at $\beta$ $=$ 0 and decreases as $\beta$ increases towards both the prolate and oblate sides indicating quenching due to deformation. Also, DF in spherical nucleus $^{46}$Ar is higher (DF $=$ 0.5) ) than in deformed nucleus $^{44}$S (DF $\approx$ 0.37 at $\beta$ $=$ 0.27). This shows weakening of bubble effect in deformed nucleus which could be due to the lowering of some of the single particle states due to deformation. The occupancy of 2s$_{1/2}$ state for $^{44}$S and $^{46}$Ar shown in Fig. \ref{fig8}(b)  appears to be similar which indicates that it is not playing a major role in the quenching of bubble effect in deformed nuclei. Moreover, it may be noted that the oblate states are fully occupied whereas the prolate states have almost zero occupancy. But the depletion fraction is decreasing with increasing $\beta$ on prolate side even though the 2s$_{1/2}$ states is empty. This shows that although the unoccupancy of s-orbit is an important condition for central depletion, the deformation has some additional influence because of which the depletion fraction decreases on the prolate side even though the 2s-orbit is vacant. Both the parameters TMA and DD-ME2 match well in predicting the strong role of deformation in the weakening of bubble effect. \par
\begin{table}
\caption{Comparing quadrupole deformation $\beta$ with expt.~\cite{nndc} and other theories~\cite{nndc,Moller,Goriely,Horiuchi}.}
\centering
\resizebox{0.3\textwidth}{!}{%
{\begin{tabular}{|c|c|c|c|c|}
 \hline
 \multicolumn{1}{|c|}{Data}&
 \multicolumn{4}{|c|}{Nuclei}\\
\hline
&$^{24}$Ne&$^{40}$Mg&$^{44}$S&$^{32}$Ar\\
\hline
Expt. \cite{nndc}&0.41&-&0.25&0.26\\
RMF&0.22&0.45&0.27&0.21\\
NSM&0.23&0.29&0.16&0.20\\
FRDM \cite{Moller}&0.06&0.31&0.25&0.28\\
HFB \cite{Goriely}&0.24&0.31&0.27&0.21\\
SKM* \cite{Horiuchi}&0.20&-  &0.10 &-  \\
\hline
\end{tabular}}}
\end{table}
Another interesting observation is to notice that the DF on oblate side is decreasing more rapidly than on the prolate side. This rapid decline in DF for oblate deformation may be attributed to the combined effect of deformation and the full occupancy of 2s$_{1/2}$ state on the oblate side. To probe this further, we plot DF vs. deformation for Si isotopes in Fig. \ref{fig8}(c). The deformations in the Si isotopes vary from many oblate to few spherical and few prolate. The depletion fraction for spherical nuclei $^{34,36,48}$Si is the highest (0.25$-$0.3) while in the deformed nuclei, the DF is usually lower in nuclei with oblate deformation than for nuclei with prolate deformation even with similar $\beta$ value. This indicates the oblate deformation to have smaller DF than the prolate deformation. However, we need to probe further to get more clarity on this aspect.\par

\begin{figure}[htb]
\centering
\includegraphics[width=8cm,height=10cm,keepaspectratio]{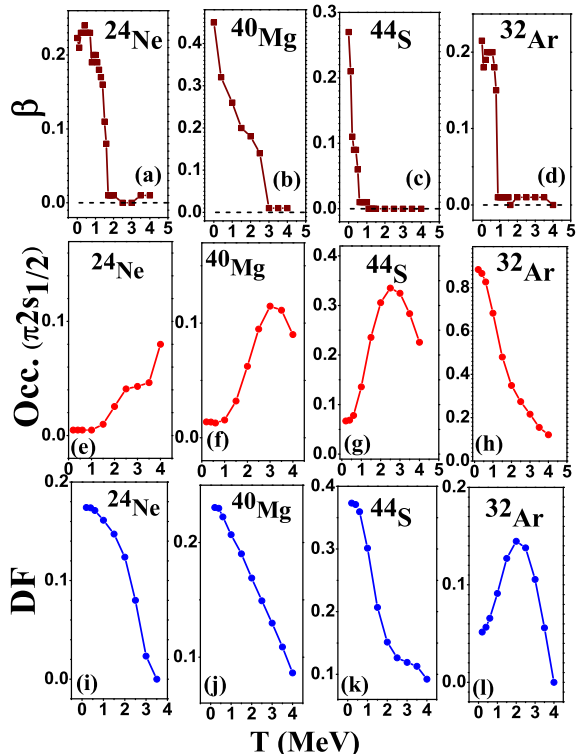}
\caption{(Colour online) (a)-(d) deformation $\beta$, (e)-(h) occupation probability, and (i)-(l) DF for $^{24}$Ne, $^{40}$Mg, $^{44}$S and $^{32}$Ar vs. T.}
\label{fig9}
\end{figure}
Since the long range correlations and dynamical quadrupole shape effects~\cite{wu} quench but not eliminate the bubble effect~\cite{yao,yao1,grasso,li} whereas the temperature eliminates it at a critical temperature T$_c$. Therefore, we evaluate and plot ((a)-(d)) the deformation, ((e)-(h)) occupancy of 2s-orbit and ((i)-(l)) the depletion fraction,  vs. T $=$ 0 to 4 MeV as shown in Fig. \ref{fig9}. At T $=$ 0, the equilibrium deformation is large with $\beta$ $\approx$ 0.2$-$0.4, the occupancy in 2s$_{1/2}$ state is zero and DF shows the highest value corresponding to each nucleus indicating bubble effect in deformed $^{24}$Ne, $^{40}$Mg and $^{44}$S except in $^{32}$Ar (to be discussed later). With increasing T, the deformation reduces to almost zero as expected~\cite{MAPRC80,MAPRC69} at around T $=$ 1$-$1.5 MeV, except  for $^{40}$Mg which remains deformed upto T $=$ 3 MeV. The occupancy in 2s$_{1/2}$ increases with increasing T and DF decreases. As is reflected from Fig. \ref{fig9} that at low T, the depletion fraction decreases as a function of T very slowly initially under the combined effect of low temperature and relatively high deformation. Also, the decline in DF in $^{40}$Mg is very slow upto T $=$ 3 MeV till it is deformed, as compared to that of $^{24}$Ne and $^{44}$S  where deformation has vanished at T $\approx$ 1 MeV after which the decline in DF with T is much more rapid. Here it should be noted that since the deformation has quenching effect on the central depletion, the quenching of bubble effect should decrease with the decreasing deformation and as a result DF should increase. But here, DF is decreasing even though the deformation is decreasing. This is because the temperature is playing a predominant role in dampening the shell and deformation effects and consequently reducing the deformation as well as the central depletion which reflects in the lowering of DF. Once the deformation vanishes, the quenching of bubble effect is attributed to temperature and then the bubble effect vanishes at a critical temperature the way it does in spherical nuclei. This shows a strong correlation between the deformation, the occupancy in 2s-orbit and the depletion fraction. Here, it is important to mention that the our calculations using state dependent BCS method show that the pairing decreases monotonically as temperature increases with pairing collapse at T $=$ 2 MeV for $^{32}$Ar, and T $=$ 3 MeV for $^{24}$Ne and $^{44}$S in contrast to pairing collapses at T $\approx$ 0.5 MeV seen in Refs. \cite{TAN,tagrawal,goodman}. Since pairing correlations also quench the depletion and therefore, one would expect an increase in the value of DF with decreasing pairing correlation, but DF decreases as the temperature increases for all the nuclei considered irrespective of the variation of pairing. Therefore as temperature increases, it becomes a predominant factor responsible for the quenching of bubble even if zero pairing and zero deformation favour bubble.\par

The variation of occupancy and DF of proton rich $^{32}$Ar appears to be different from that of the other deformed nuclei. In Figs. \ref{fig9} ((d), (h) and (l)), the occupancy is high around 0.8 and DF has a very low value at T $=$ 0. Thus $^{32}$Ar does not qualify to be a bubble candidate which contradicts the prediction of Ref.~\cite{shukla}. The reason for the weakening of central depletion could be due to the shell effects and structural transitions in $^{32}$Ar. The inversion of 2s$_{1/2}$ and 1d$_{3/2}$ state~\cite{todd,grasso}, which leads to the bubble effect in neutron rich $^{46}$Ar (seen in Fig. \ref{fig6} and~\cite{plb2018}) is not observed in neutron deficient $^{32}$Ar. The 2s$_{1/2}$ state lies just below the Nilsson 1d$_{3/2}$ level with full occupancy which leads to smaller DF in $^{32}$Ar. As T increases, the occupancy in 2s-orbit decreases due to the particles occupying higher levels and hence DF increases. Relatively high value of DF at T $\approx$ 2 MeV with zero deformation and small occupancy is indicative of central depletion although it may not be very pronounced.  For T $>$ 2 MeV, DF starts decreasing which indicates the usual quenching of bubble effect as seen (in Fig. \ref{fig9}) in other deformed nuclei.  With T increasing further, shell effects are washed out, deformation reduces to zero and the occupancy of 2s-state becomes high and consequently the bubble effect also vanishes. Hence, at low temperature, the shell structure is important and the bubble structure is characterized by the (un)occupancy of 2s$_{1/2}$ state along with the deformation in case of deformed nuclei. \par

To conclude, deformation and temperature induced effects on the bubble nuclei are studied using the temperature dependent RMF plus state dependent BCS theory in the light, medium, heavy and superheavy nuclei. The quenching effect of deformation on the deformed bubble candidates is studied at temperature T $\ge$ 0. At T $=$ 0, the well deformed nuclei $^{24}$Ne, $^{40}$Mg and $^{44}$S  exhibit central depletion which is significantly influenced by deformation. DF decreases for the prolate deformation even if the occupancy of s-orbit is zero which shows the prominent role of deformation apart from the depopulation of 2s-orbit as usually believed. The oblate deformation shows smaller DF than the prolate deformation. However, this needs further investigation.\par
As the temperature increases, the shell effects start melting away, the deformation vanishes slowly, the occupancy of s-orbit near Fermi level increases and as a result, the DF decreases. The bubble structure gets completely erased at a critical temperature T$_c$ $\approx$ 3$-$5 MeV. The density distribution, single particle spectra, occupation probability and charge form factor  demonstrate a bubble at T $=$ 0 and a non-bubble at T $>$ 3 MeV which establishes the anti-bubble effect of temperature. The variation of occupation probability of 2s-orbit with T agrees with that from statistical theory of hot nuclei.\par
In superheavy nuclei, the effect of temperature on central depletion is not as significant as in the lighter nuclei. With increasing T, the central depletion decreases to some extent due to melting of shell effects but does not vanish as it does in light nuclei at T$_c$. The central depletion sustains due to the combined effect of large repulsive Coulomb forces and the attractive nucleon forces and the shell effects play only a subtle role. Since there are not many studies available on this subject, many more efforts on the experimental as well as theoretical fronts using the other theoretical models are very much needed to get much more clarity on this subject. \par
G.S. and M.A. acknowledge the support provided by SERB (DST), Govt. of India under YSS/2015/000952 and WOS-A scheme respectively. We are specially thankful to Prof. M. Wakasugi, Team Leader, SCRIT Team for sharing experimental inputs and Prof. B. K. Agrawal, SINP, India for the discussions.


\begin{thebibliography}{99}
\bibitem{nature} A. Mutschler \textit{et al.}, Nature Physics 13 (2017) 152.
\bibitem{campi}X. Campi and D.W.L. Sprung, Phys. Lett. B 46 (1973) 291.
\bibitem{todd} B.G. Todd-Rudel, J. Piekarewicz, and P.D. Cottle, Phys. Rev. C 69 (2004) 021301R.
\bibitem{grasso} M. Grasso, Z. Ma, E. Khan, J. Margueron, and Nguyen Van Giai, Phys. Rev. C 76 (2007) 044319.
\bibitem{khan} E. Khan, M. Grasso, J. Margueron, and N. Van Giai, Nucl. Phys. A 800 (2008) 37.
\bibitem{wang} Y.Z. Wang, J.Z. Gu, X.Z. Zhang, and J.M. Dong, Chin. Phys. Lett. 28 (2011) 102101.
\bibitem{wang1} Y.Z. Wang, J.Z. Gu, X.Z. Zhang, and J.M. Dong, Phys. Rev. C 84 (2011) 044333.
\bibitem{grasso1} M. Grasso \textit{et al.}, Phys. Rev. C 79 (2009) 034318.
\bibitem{yao1} J.M. Yao, S. Baroni, M. Bender, and P.-H. Heenen, Phys. Rev. C 86 (2012) 014310.
\bibitem{schuetrumpf} B. Schuetrumpf, W. Nazarewicz, and P.-G. Reinhard, Phys. Rev. C 96 (2017) 024306.
\bibitem{sobi} A. Sobiczewski and K. Pomorski, Prog. Part. and Nucl. Phys. 58 (2007) 292.
\bibitem{decharge} J. Decharge, J.-F. Berger, K. Dietrich, and M.S. Weiss, Phys. Lett. B 451 (1999) 275.
\bibitem{sksingh} S.K. Singh, M. Ikram, and S.K. Patra, Int. J. Mod. Phys. E 22 (2012) 135001.
\bibitem{ikram} M. Ikram, S.K. Singh, A.A. Usmani, and S.K. Patra, Int. J. Mod. Phys. E 23 (2014) 1450052.
\bibitem{bender} M. Bender and P.-H. Heenen, J. Phys. Conf. Ser. 420 (2013) 012002.
\bibitem{wu} X.Y. Wu, J.M. Yao and Z.P. Li, Phys. Rev. C  89 (2014) 017304.
\bibitem{yao} J.M. Yao, H.Mei, Z.P. Li., Phys. Lett. B 723 (2013) 459.
\bibitem{li} J.J. Li, W.H. Long, J.L. Song, and Q. Zhao, Phys. Rev. C 93 (2016) 054312.
\bibitem{duguet} T. Duguet, V. Soma, S. Lecluse, C. Barbieri, and P. Navratil, Phys. Rev. C 95 (2017) 034319.
\bibitem{plb2018} G. Saxena, M. Kumawat, M. Kaushik, S.K. Jain, and Mamta Aggarwal, Accepted in Phys. Lett. B (2018).
\bibitem{TAN} L. Tan Phuc, N. Quang Hung, and N. Dinh Dang, Phys. Rev. C 97 (2018) 024331.
\bibitem{nakada} H. Nakada, K. Sugiura, and J. Margueron, Phys. Rev. C 87 (2013) 067305.
\bibitem{BKbubble} G. Saxena, M. Kumawat, B.K. Agrawal, and Mamta Aggarwal, Communicated to J. Phys. G: Nucl. Part. Phys. (2018).
\bibitem{MAPLB} Mamta Aggarwal, Phys. Lett. B 693 (2010) 489.
\bibitem{MAPRC} Mamta Aggarwal, Phys. Rev. C 90 (2014) 064322.
\bibitem{MAPRC69} Mamta Aggarwal, Phys. Rev. C 69 (2004) 034602.
\bibitem{MAPRC80} Mamta Aggarwal and I. Mazumdar, Phys. Rev. C 80 (2009) 024322.
\bibitem{saxena} G. Saxena, M. Kumawat, M. Kaushik, U.K. Singh, S.K Jain, S. Somorendro Singh, and Mamta Aggarwal, Int. J. Mod. Phys. E 26 (2017) 1750072.
\bibitem{saxena3} G. Saxena, M. Kumawat, M. Kaushik, S.K. Jain, and Mamta Aggarwal, Phys. Lett. B 775 (2017) 126.
\bibitem{suga} Y. Sugahara and H. Toki, Nucl. Phys. A 579 (1994) 557.
\bibitem{Lalazissis09} G.A. Lalazissis, S. Karatzikos, R. Fossion, D.P. Arteaga, A. Afanasjev, and P. Ring, Phys. Lett. B 671 (2009) 36.
\bibitem{Lalazissis05} G.A. Lalazissis, T. Niksic, D. Vretenar, and P. Ring, Phys. Rev. C 71 (2005) 024312.
\bibitem{gambhir} Y.K. Gambhir, P. Ring, and A. Thimet; Ann. Phys. (N.Y.) 198 (1990) 132.
\bibitem{geng1} L.S. Geng, H. Toki, S. Sugimoto, and J. Meng, Prog. Theor. Phys. 110 (2003) 921.
\bibitem{hirata} D. Hirata, K. Sumiyoshi, I. Tanihata, Y. Sugahara, T. Tachibana and H. Toki, Nucl. Phys. A 616 (1997), 438.
\bibitem{yadav2004} H.L. Yadav, M. Kaushik and H. Toki, Int. J. Mod. Phys. E, 13 (2004) 647.
\bibitem{saxena5} G. Saxena, D. Singh, M. Kaushik, H.L. Yadav, and H. Toki, Int. Jour. Mod. Phys. E 22 (2013) 1350025.
\bibitem{Dobaczewski1995} J. Dobaczewski, W. Nazarewicz, T.R. Werner, J.F. Berger, C.R. Chinn and J. Decharge, Phys. Rev. C 53 (1996) 2809.
\bibitem{Goriely2002} S. Goriely, M. Samyn, P.H. Heenen, J.M. Pearson and F. Tondeur, Phys. Rev. C 66 (2002) 024326.
\bibitem{saxena4} D. Singh, G. Saxena, M. Kaushik, H.L. Yadav and H. Toki, Int. Jour. Mod. Phys. E 21 (2012) 1250076.
\bibitem{tbrack} M. Brack and P. Quentin, Phys. Scr. A10, 163 (1974).
\bibitem{tquentin} P. Quentin and H. Flocard, Annu. Rev. Nucl. Part. Sc. 28, 523 (1978).
\bibitem{tgambhir} Y. K. Gambir, J. P. Maharana, G. A. Lalazissis, P. Panos and P. Ring, Phys. Rev. C 62 (2000) 054610.
\bibitem{tagrawal} B. K. Agrawal, Tapas Sil, J. N. De and S. K. Samaddar, Phys. Rev. C 62 (2000) 044307.
\bibitem{ttapas} Tapas Sil, B. K. Agrawal, J. N. De, and S. K. Samaddar, Phys. Rev. C 63 (2001) 064302.
\bibitem{tsenthil} M.T. Senthil Kannan, Bharat Kumar, M. Balasubramaniam, B. K. Agrawal, S. K. Patra Phys. Rev. C 95 (2017) 064613.
\bibitem{tbharat} Bharat Kumar, M. T. Senthil Kannan, M. Balasubramaniam, B. K. Agrawal, and S. K. Patra, Phys. Rev. C 96 (2017) 034623.
\bibitem{goodman} A. L. Goodman, Nucl. Phys. A 352 (1981) 30.
\bibitem{quang} N. Quang Hung and N. Dinh Dang, Phys. Rev. C 79  (2009) 054328.
\bibitem{dinh} N. Dinh Dang and N. Quang Hung, Phys. Rev. C 77 (2008) 064315.
\bibitem{moretto} L. G. Moretto, Phys. Lett. B 40 (1972) 1.
\bibitem{dinh1} N. Dinh Dang, P. Ring, and R. Rossignoli, Phys. Rev. C 47 (1993) 606.
\bibitem{dinh2} N. Dinh Dang and A. Arima, Phys. Rev. C 68 (2003) 014318.
\bibitem{suda} T. Suda and M. Wakasugi, Prog. Part. Nucl. Phys. 55 (2005) 417.
\bibitem{suda1} T. Suda \textit{et al}., Phys. Rev. Lett. 102 (2009) 102501.
\bibitem{hofstadter} R. Hofstadter, Rev. Mod. Phys. 28 (1956) 214.
\bibitem{forest} T. de Forest Jr. and J.D. Walecka, Adv. Phys. 15 (1996) 1.
\bibitem{donnelly} T. W. Donnelly and J.D. Walecka, Annu. Rev. Nucl. Part. Sci. 25 (1975) 329.
\bibitem{meucci} A. Meucci, M. Vorabbi, C. Giusti, F.D. Pacati, and P. Finelli, Phys Rev. C 89 (2014) 034604.
\bibitem{shukla} A. Shukla and S. Aberg, Phys. Rev. C 89 (2014) 014329.
\bibitem{MAMPRC89} M. Aggarwal, Phys. Rev. C 89 (2014) 024325.
\bibitem{nndc} http://www.nndc.bnl.gov/chart/.
\bibitem{Moller} P. Moller, A.J. Sierk, T. Ichikawa, and H. Sagawa, At. Data Nucl. Data Tables 109-110 (2016) 1.
\bibitem{Goriely}S. Goriely \textit{et al}., Phys. Rev. Lett. 102 (2009) 152503, http://www-astro.ulb.ac.be/bruslib.
\bibitem{Horiuchi} W. Horiuchi \textit{et al}., Phys. Rev. C 86 (2012) 024614.

\end{thebibliography}
\end{document}